\documentstyle[sprocl]{article}

\input{psfig.sty}

\def\be{\begin{equation}}
\def\ee{\end{equation}}
\def\bea{\begin{eqnarray}}
\def\eea{\end{eqnarray}}
\def\lesssim{\mathrel{\raise1.16pt\hbox{$<$}\kern-7.0pt 
\lower3.06pt\hbox{{$\scriptstyle \sim$}}}}         

\begin{document}


\title{INTRINSIC GALAXY ALIGNMENTS AND WEAK GRAVITATIONAL LENSING}

\author{A. F. HEAVENS}

\address{Institute for Astronomy, University of Edinburgh\\Blackford Hill,
Edinburgh EH9 3HJ, UK\\E-mail: afh@roe.ac.uk}


\maketitle\abstracts{Gravitational lensing causes background
galaxy images to become aligned, and the statistical characteristics
of the image alignments can then be used to constrain the power
spectrum of mass fluctuations.  Analyses of gravitational lensing
assume that intrinsic galaxy alignments are negligible, but if
this assumption does not hold, then the interpretation of image
alignments will be in error.  As gravitational lensing
experiments become more ambitious and seek very low-level
alignments arising from lensing by large-scale structure, it
becomes more important to estimate the level of intrinsic
alignment in the galaxy population.  In this article, I review
the cluster of independent theoretical studies of this issue, as
well as the current observational status.  
Theoretically, the calculation of intrinsic alignments is by no means
straightforward, but some consensus has emerged from the existing
works, despite each making very different assumptions. This consensus
is that\\
\indent  a) intrinsic alignments are a small but non-negligible
($\lesssim 10 \%$) contaminant of the lensing ellipticity correlation
function, for samples with a median redshift $\bar z \sim 1$;
\\
\indent b) intrinsic alignments dominate the signal for
low-redshift samples ($\bar z \sim 0.1$), as expected in the
SuperCOSMOS lensing survey and the Sloan Digital Sky Survey.}

\section{Introduction}
\subsection{Lensing and galaxy shapes}
Galaxy shapes are
being used increasingly as a tool in cosmology. Specifically,
correlations of the orientations of galaxies on the sky, which
arise as a result of gravitational lensing, can be used to
measure the mass distribution in the Universe.  As cosmology
becomes more of a precision science, with accurate estimation of
cosmological parameters possible from microwave background
observations, weak lensing offers a complementary accurate method
of studying the state of the cosmos. Traditional studies of
large-scale structure focus on the clustering properties of
galaxies, but this approach suffers from the disadvantage that
galaxies may not be unbiased tracers of the mass distribution.
Since it is only the mass distribution which is robustly
predicted from theoretical models, it is very attractive to
pursue methods which measure this distribution rather directly,
without having to make assumptions about where galaxies should
form in a given theoretical mass distribution.

Gravitational lensing is now established as a powerful method to
measure directly the distribution of mass in the universe (e.g. Gunn
1967; Bartelmann \& Schneider 1999 and references therein). This
method is based on the measurement of the coherent distortions that
lensing induces on the observed shapes of background galaxies.
Recently, several groups have reported the statistical detection of
weak lensing by large-scale structure (Wittman et al. 2000; van
Waerbeke et al. 2000; Bacon, Refregier \& Ellis 2000; Kaiser, Wilson
\& Luppino 2000). These detections offer remarkable prospects for
precise measurements of the mass power spectrum and of cosmological
parameters (e.g. Hu \& Tegmark 1998).

An assumption which is made in the interpretation of alignments in
galaxy images is that the directions of projected images are
intrinsically uncorrelated.  At first sight, this seems a reasonable
assumption to make; the redshift distribution of the background
galaxies is usually very broad, so most source galaxies are not
physically close.  However, the signal which is sought in weak lensing
studies is very small, with typical ellipticity correlations of $\sim
10^{-4}$.  It is by no means obvious that the intrinsic correlations
will be small enough to be negligible for lensing studies.

There are sound reasons for expecting some correlations in galaxy
shapes.  The shapes may be determined in part by the tidal
gravitational field, which will have correlations at some level, or by
correlation of torques during linear evolution. Shapes may also be
influenced by merger events, which may not be isotropically
distributed, but influenced by filamentary or other structures.

The realisation that intrinsic galaxy alignments might be
important was made by a number of groups simultaneously,
resulting in a number of independent studies (Heavens, Refregier
\& Heymans 2000, Pen, Lee \& Seljak 2000, Croft \& Metzler 2000,
Crittenden, Natarajan, Pen \& Theuns 2000, Catalan, Kamionkowski
\& Blandford 2001).  It should be said that a calculation of intrinsic
shape correlations is by no means trivial. Analytically, the problem
is very difficult; numerically, it is hard to get sufficient
signal-to-noise; finally, it is plausible that shapes and their
correlations are at least partly determined by non-gravitational
processes which are difficult to model.  In summary we are far from
being able to predict robustly the ellipticity distribution of
galaxies.

In order to make progress, the studies have made very different
simplifying assumptions in relating the density field to the
ellipticities of galaxies.  The remarkable outcome is,
however, a level of agreement of the level of the effect which
one would perhaps not have expected in advance.   The conclusion
of all the studies is that for a broad distribution of sources
centred around $z=1$, the alignment intrinsic in the galaxy
distribution is a small but not insignificant contributor to the
total expected from lensing and intrinsic effects.  For a
low-redshift ($z \simeq 0.1$) source population, where the
intrinsic effects are larger and the lensing effect smaller, the
alignment of images is dominated almost completely by intrinsic
effects.
\begin{figure}[t]
\psfig{figure=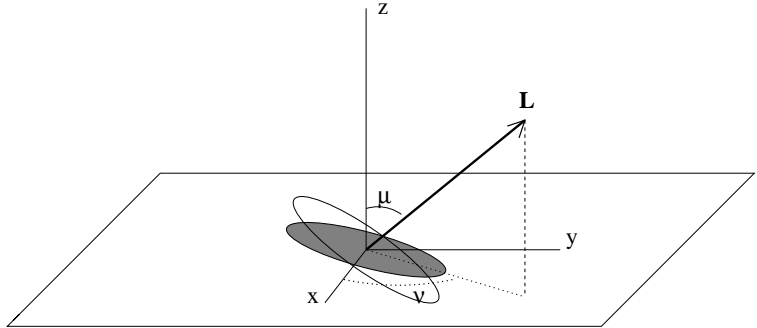,height=1.7in} \caption{Simple model of a
spiral galaxy. The disk of the galaxy is shown as the open
ellipse, and its projection on the sky as the filled
ellipse.\label{spiraldiagram}}
\end{figure}

\section{Assumptions: galaxy ellipticity}

Catelan et al (2000) assumed that the shape is determined by the
linear tidal field; Mackey et al. (2001) extended this model to
consider disk galaxies in more detail.  Lee, Pen \& Seljak (2000) and
Crittenden et al. (2000) used a correlation, seen in N-body
simulations between the tidal field and the moment of inertia tensor
(Lee \& Pen 2000).  From this they determined the angular momentum
vector distribution.  Numerical simulations from the VIRGO consortium
were analysed by Heavens et al (2000) and Croft \& Metzler (2000),
with similar conclusions.  Both groups investigated an `elliptical'
model, where the ellipticity of the visible galaxy and the halo were
assumed equal.  Heavens et al. also looked a `spiral' model, assuming
that the galaxy was a thin disk with angular momentum vector aligned
with that of the halo.  Given that one is seeking a very small
correlation of ellipticities, it would not be very surprising if these
different assumptions led to rather different answers.  This turns out
not to be the case.
\begin{figure}[t]
\psfig{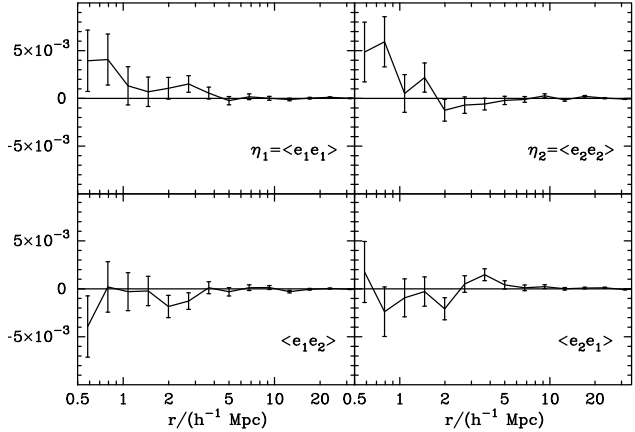} 
\caption{Three-dimensional
correlation functions for `spirals' in a $\Lambda CDM$ model at a
redshift $z=1$.\label{Cspi}}
\end{figure}

\section{Numerical simulations}

I will describe here the two approaches which use numerical
simulations to estimate galaxy ellipticity correlations.  The articles
by Rob Crittenden and Jonathan Mackey in this volume will concentrate
on more analytic approaches. Both numerical papers use outputs from
Virgo simulations of 17 million particles, from which haloes are
extracted using a friends-of-friends algorithm.  Ellipticities in the
spiral model are determined by the orientation of the angular momentum
vector of the halo ${\mathbf L}$ with the line of sight (see
fig. \ref{spiraldiagram}). With the coordinate system shown in the
figure, the ellipticity $e \equiv ( e_{1}^{2} + e_{2}^{2}
)^{\frac{1}{2}}$ and position angle $\alpha$ of the projected ellipse
are given by
\begin{eqnarray}
\label{eq:disk}
e & = & \frac{\sin^{2} \mu}{1+\cos^{2} \mu}, \nonumber \\
\alpha & = & \nu + \frac{\pi}{2}.
\end{eqnarray}
and the ellipticity is $e_{i} = e\{ \cos 2 \alpha, \sin 2 \alpha \}$.
Note that the observed ellipticity depends only on the orientation of
${\mathbf L}$, and not on its magnitude; it is also independent of the
surface brightness profile, provided it depends only on radius.  In
some sense, $e$ should be taken as an extreme: the average ellipticity
of a randomly-oriented distribution of spirals is 0.57 - considerably
larger than that observed.  The ellipticity of the `elliptical' models
are calculated from halo quadrupole moments (Kaiser \& Squires 1993).
In practice, the coordinate system is rotated so that the $x-$axis
lies along the projected separation of the galaxy pairs.

\begin{figure}[t]
\psfig{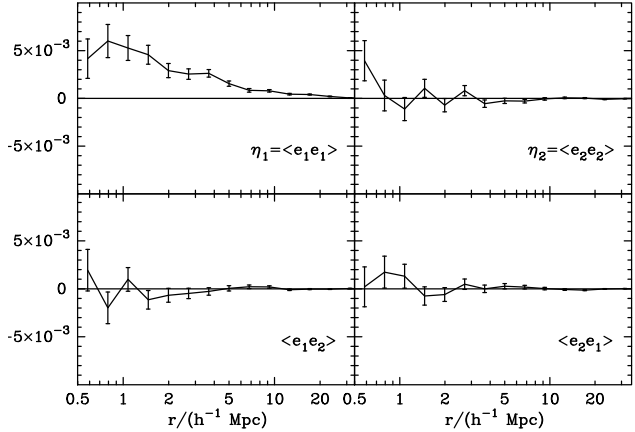} 
\caption{As fig. \ref{Cspi}, but for `ellipticals'.\label{Cell}}
\end{figure}

It turns out that the ellipticities based on the spiral and elliptical
prescriptions are very different.  The ellipticity correlation
functions, however, are rather similar (fig. \ref{Cell} and
\ref{Cspi}).  The ellipticities themselves are typically not much
less than unity, so the correlation level of $\lesssim 5
\times 10^{-3}$ is rather small;  the agreement, therefore, is by
no means expected.

\begin{figure}[t]
\psfig{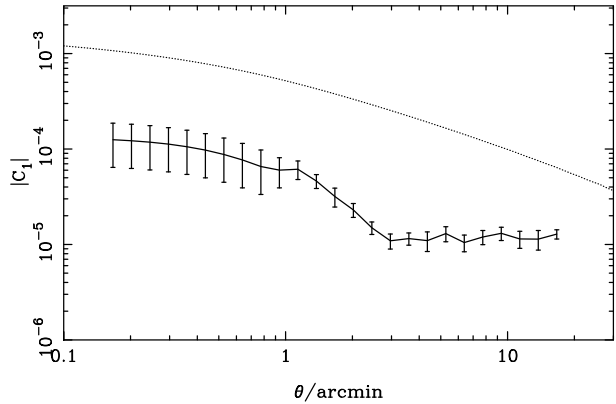} \caption{Intrinsic
ellipticity correlation function of `ellipticals', for a deep survey
with median source redshift of 1, and a broad source distribution.
Also shown dotted is the expected weak lensing correlation signal.  We see
that the lensing signal dominates for this redshift, and intrinsic
effects act as a $\sim 10\%$ contaminant.\label{Ctheta}}
\end{figure}

To assess the impact on the lensing signal, the three-dimensional
correlation functions are turned into two-dimensional correlations,
using a modified form of Limber's equation.  Fig. \ref{Ctheta} is
typical of the results obtained for a deep survey at median redshift
of unity.  We see that the expected weak lensing signal (dotted)
dominates, but the intrinsic signal is not entirely negligible,
contributing $\sim 10\%$ of the correlation.  This general conclusion
has been reached by all studies.

\begin{figure}[t]
\psfig{figure=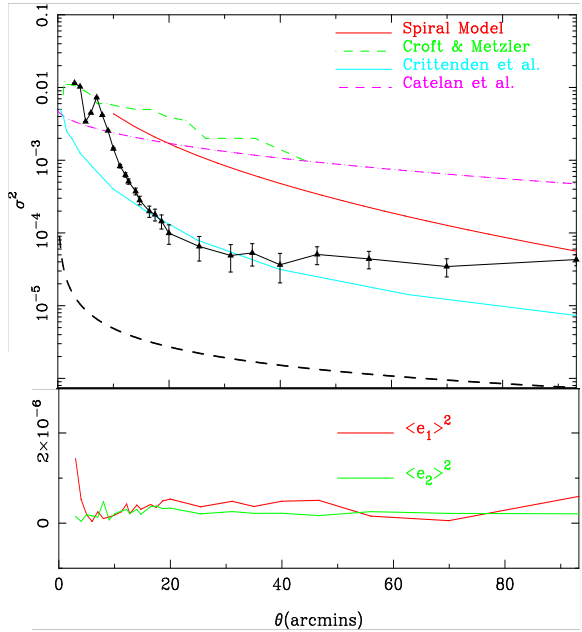,height=3.5in,angle=270} \caption{Shear variance
estimate from the SuperCOSMOS survey (Brown et al 2001), compared with
the `spiral' model of Heavens et al (dark solid line), the elliptical
model of Croft \& Metzler (upper dashed), the semi-analytic model of
Crittenden et al (fainter solid), and the tidal stretch model of Catelan
et al (lower dot-dashed). The expected lensing signal is the lowest
line.\label{Brown}}
\end{figure}

At low redshift, the story is quite different.  The lensing signal is
lower as there is less mass to pass through, and the intrinsic signal
is larger, because fixed angular separations translate to smaller
typical physical separations. For median redshifts of order unity, as
expected in SuperCOSMOS and Sloan (Gunn et al 1995), the intrinsic
effect dominates the lensing correlation by up to several orders of
magnitude.  This is illustrated in fig.
\ref{Brown}, which shows the measured shear variance (which is
related to the ellipticity correlation function) from the
SuperCOSMOS survey, along with predictions from the various
intrinsic studies (see also Pen, Lee \& Seljak 2000). Although
there are certainly some significant differences between the
studies, all agree that on small scales the intrinsic signal
should be $\sim 10^{-2}$.

\section{Discussion}

I have summarised here the state of the investigations into the level
of intrinsic alignment of galaxies, with a view to assessing its
impact on weak gravitational lensing studies.  Despite the
difficulties of computing the correlation of ellipticities with
confidence, several independent studies have come up with broadly
similar conclusions.  These are that the level of correlation is
unlikely to be low enough to be ignored.  For high-redshift ($z\simeq
1$) lensing samples, the intrinsic correlation, being diluted over a
wide redshift range, is a small, but non-negligible ($\sim 10\%$)
fraction of the correlation induced by weak lensing.  At low redshifts
($z\simeq 0.1$), however, the lensing signal is smaller and the
intrinsic correlation larger, to the extent that the intrinsic
correlation dominates the lensing signal.  This is also supported by
observed correlations of galaxy shapes from the SuperCOSMOS survey
(Brown et al. 2001), which shows distortions at a level far above the
expected signal but broadly in line with theoretical predictions.
There are some differences in the predictions for intrinsic
correlations, but all agree with the above statements.  It is perhaps
difficult to see where the theoretical predictions could be made
significantly more robust; certainly the methods based on numerical
simulations would benefit by larger number of particles in each halo,
and by larger volumes to improve signal-to-noise.  Analytic methods
are likely to be limited by the validity or otherwise of the
simplifying assumptions, and neither general method can really be
expected to include non-gravitational processes reliably in the near
future.  Thus it is probably best to be satisfied that we know the
magnitude of the effect, and therefore not attempt to do
straightforward weak lensing calculations from low-redshift samples.
One can try to eliminate partly the intrinsic correlation by
performing a curl-gradient decomposition of the ellipticity field,
since weak lensing induces only a gradient field (Crittenden et al
2001, but see Mackey et al 2001).  Alternatively, one can sidestep the
problem almost entirely; this can be done, albeit by sacrificing some
signal-to-noise, by cross-correlating orientations of galaxies which
are known to be distant physically, for example by using photometric
redshift information.

\section*{Acknowledgments}
I am grateful to my collaborators, Alexandre Refregier and Catherine
Heymans, and also to Andy Taylor for the figures from the SuperCOSMOS
survey.  The simulations analysed in this paper were carried out using
data made available by the Virgo Supercomputing Consortium
(http://star-www.dur.ac.uk/~frazerp/virgo/) using computers based at
the Computing Centre of the Max-Planck Society in Garching and at the
Edinburgh Parallel Computing Centre.  We are very grateful to Rob
Smith for providing halos from the simulations.

\end{document}